%% file: p-Conf-99-201-E.tex
\documentstyle[preprint,epsfig,aps,amsmath,floats,tighten]{revtex} 

\def\pbar{$\overline{p}~$}               
\def\tbar{$\overline{t}~$}               
\def\pbarp{$\overline{p}p~$}             
\def\qqbar{$q\overline{q}~$}             
\def\pt{$p_T~$}                          
\def\met{\mbox{${\hbox{$E$\kern-0.6em\lower-.1ex\hbox{/}}}_T~$}} 
\def\D0{D\O}                            
\begin{document}
\lefthyphenmin=2
\righthyphenmin=3

%
%
\title{
Measurement of the Angular Distribution of Electrons from $W
\to e \nu$ Decays Observed in $p \bar p$ Collisions at $\sqrt{s}= 1.8$~TeV
}

\author{\centerline{The D\O\ Collaboration
  \thanks{Submitted to the {\it International Europhysics Conference} on
         {\it High Energy Physics}, {\it EPS-HEP99},
          \hfill\break
           15 -- 21 July, 1999, Tampere, Finland.}}}
\address{
\centerline{Fermi National Accelerator Laboratory, Batavia, Illinois 60510}
}

%
%
\date{\today}

\maketitle

%
%
\begin{abstract}
We present a preliminary measurement of the electron angular
distribution parameter ${\alpha_{2}}$
in $W \to e \nu$ events using data collected by the {D\O} detector
during the 1994--1995 Tevatron run. We compare our results with
next-to-leading order perturbative QCD, which predicts an angular
distribution of ($1 \pm \alpha_1 \, \mathrm{cos} \theta^* +
\alpha_2\,  \mathrm{cos}^2 \theta^*$), where $\theta^*$ is the
angle between the charged lepton and the antiproton in the
Collins-Soper frame.  In the presence of QCD corrections, the
parameters $\alpha_1$ and $\alpha_2$ become functions of $p_T^W$, the
$W$ boson transverse momentum.  We present the first measurement of
$\alpha_2$ as a function of $p_T^W$. This measurement is of
importance, because it provides a test of next-to-leading order QCD
corrections which are a non-negligible contribution to the $W$ mass
measurement.
\end{abstract}

\newpage
\begin{center}
\input{list_of_authors}
\end{center}

\vfill\eject

\section{Introduction}
\label{sec:intro}
At the Fermilab Tevatron, which operates at a
center of mass energy of {\mbox{$\sqrt{s}$ =\ 1.8\ TeV}},
$W$ and $Z$ bosons
are produced in high energy \pbarp collisions.
In addition to probing electroweak physics, the study of the
production of $W$ and $Z$ bosons
provides an avenue to explore QCD, the theory of strong
interactions.  
The benefits of using intermediate vector bosons to
study perturbative QCD are large momentum transfer, distinctive
event signatures, low backgrounds, and a well understood electroweak
vertex.  
In this paper we present the measurement of the angular distribution of electrons 
from $W$ boson decays based on data taken by the \D0 collider
detector during the 1994--1995 Tevatron running period.  


In the parton model, $W$ and $Z$ intermediate vector bosons
are produced, at lowest order, in head-on collisions of \qqbar constituents of the 
proton and antiproton, and cannot have any transverse momentum.
This purely electroweak tree-level process is determined by the $(V-A)$ character of
electroweak interactions and leads to a rather simple angular dependence
of the cross section (measured by UA1~\cite{ua1cos,ua1,ua12}):
\begin{equation}
{\frac{d\sigma}{d(\cos\theta^{*})}\propto
(1\pm \cos\theta^{*})^{2}
}
\label{eqn:vma}
\end{equation}
where $\theta^*$ is the lepton angle in the Collins-Soper 
rest frame~\cite{cs} of the 
$W$ boson in which the $z$-axis is defined as the bisector of the proton 
momentum and the negative of the antiproton momentum.

\begin{figure}[!htbp]
\vbox{
\centerline{
\epsfig{figure=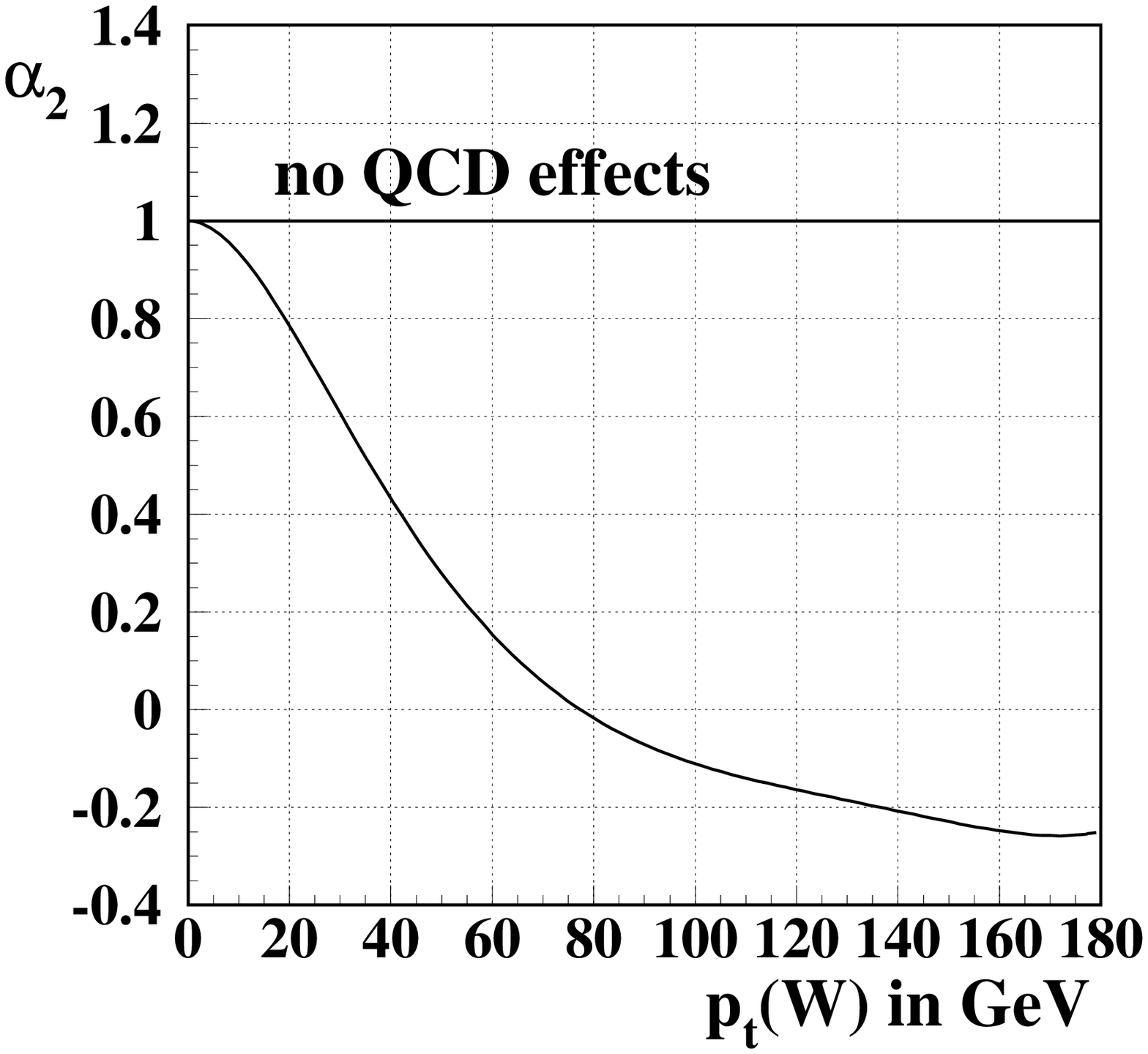,height=7cm,width=7cm}
\epsfig{figure=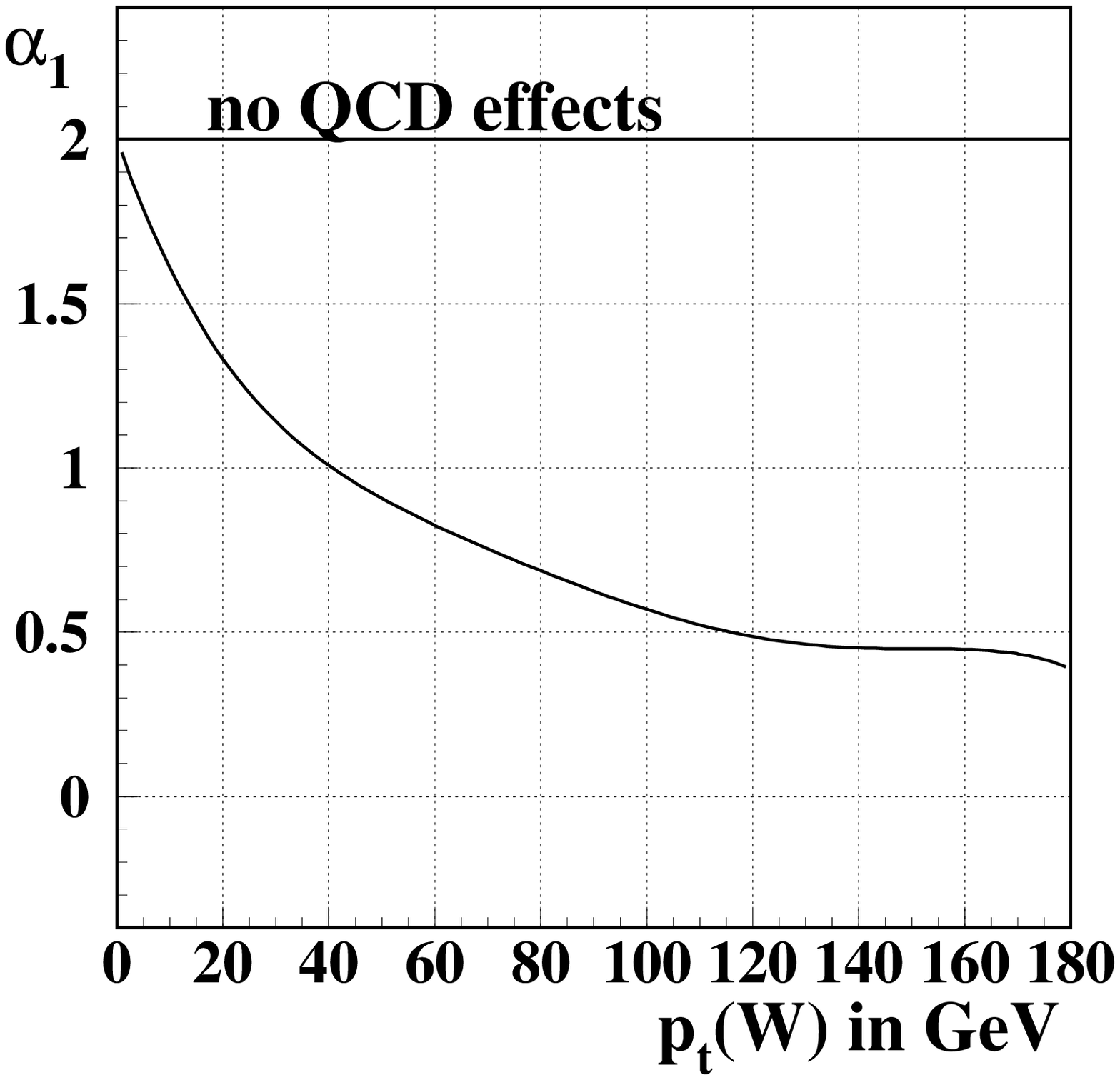,height=7cm,width=7cm}
}
\vspace{-0.0in}
\caption{The angular parameters $\alpha_{2}$ (left) and  
$\alpha_{1}$ (right) as 
a function of $p_T^W$.}
\label{fig:a2pt}
}
\end{figure}

At large transverse momentum ($p_T > 20 $ GeV), 
the cross section is dominated by the
radiation of a single parton with large transverse momentum.
Perturbative QCD~\cite{ARtheory,AKtheory} is therefore expected to be reliable in
this regime.  
The additional gluon or quark jet in high \pt\ collisions  alters 
the helicity-state of the $W$ boson; 
a calculation to next-to-leading order in QCD 
leads to nine helicity amplitudes for the various contributing processes
(see~\cite{mi}). 
The cross section can  be written in terms of the azimuthal and 
polar angles. Integrating  over the azimuthal angle leads to:
\begin{equation}
\frac{d^3\sigma}{dq_{T}^{2}\, dy\, d\cos\theta^*}=C(1+P(W)\alpha_{1}\cos\theta^*+\alpha_{2}\cos^{2}\theta^*)
\end{equation}
where $P(W)$ is the polarization of the $W$ boson. 
The angular parameters $\alpha_1$ and $\alpha_2$ are functions of the 
transverse momentum of the $W$ boson, $p_T^W$, as shown in 
figure~\ref{fig:a2pt}. In this paper we present a measurement of the 
parameter $\alpha_2$ as 
a function of $p_T^W$. 
The  angular parameter  $\alpha_1$  could not be measured by the 
\D0 detector during 
Tevatron Run 1 because the lack of a magnetic field, for  sign identification 
of electrons, makes it impossible to determine the polarization of the $W$ boson.
For  Tevatron Run II, a central solenoid magnet will be installed,
allowing for the measurement 
of both angular parameters.

The measurement of $\alpha_2$ can serve as
a probe for perturbative QCD independent of inclusive measurements. 
Since the transverse mass of the $W$ boson is correlated with the decay
angle of the lepton, QCD effects introduce a 
systematic shift to the $W$ mass measurement at \D0 , where a  
fit to the transverse mass distribution is used to determine the $W$ mass.
The shift, introduced by perturbative QCD, is $O$(40 MeV) ~\cite{note3198} for events with $p_T\leq $ 15 GeV used in the mass
measurement. In Run II, when the total error of the $W$ mass
will be reduced from the current 105 MeV~\cite{wmassprd,wmassprl} to 
an estimated 50 MeV for 1 $\rm fb^{-1}$ and  to about 30 MeV for 10 ${\rm fb}^{-1}$~\cite{TEV2000}, a good understanding of this
systematic shift is important.

\section{Experimental Method}
\subsection{Extraction of  the lepton angle \label{extraction}}
To directly measure the decay angle $\theta^*$ of the electron 
in the 
Collins-Soper frame, all momenta in the lab frame have to be
known to perform the boost to this particular rest frame of the $W$ Boson.
This is not possible, however, since the longitudinal momentum of 
the neutrino cannot be measured. 
A solution to this problem is to use the  correlation between $\cos\theta ^*$ and the transverse 
$W$ mass to infer $\cos\theta ^*$ on a statistical basis.
This is done using a Bayesian approach. 

\begin{figure}[!htbp]
\vbox{
\centerline{
\epsfig{figure=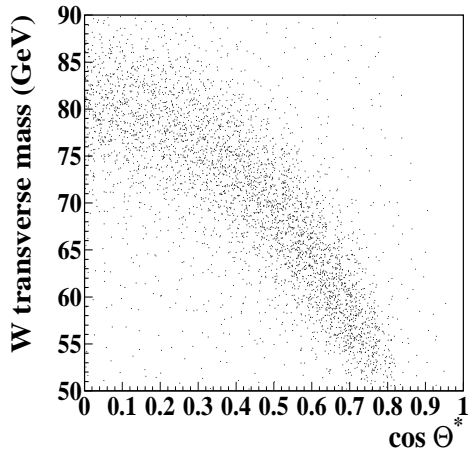,height=8cm,width=7.2cm}}
\vspace{-0.5cm}
\caption{Smeared $W$ transverse mass versus true $\cos\theta ^*$ for $p_{T}\leq$  
10 GeV. Acceptance cuts are applied to this plot. This correlation
plot is used to infer the $\cos \theta^*$ distribution from the 
measured $m_T^W$ distribution.}
\label{fig:corr}
}
\end{figure}

Figure~\ref{fig:corr} generated from  Monte Carlo events that were 
run through a parameterized detector simulation (see  
~\cite{wmassprd} ~\cite{flattum,adam,ptwprl}) 
shows the correlation of the smeared $W$ transverse  mass and the 
true value of  $\cos\theta ^{*}$. By correlating the smeared transverse 
mass as it would be measured in the \D0\ detector and the true (unsmeared)
value of $\cos\theta ^{*}$, the Bayesian analysis described below will yield
the unsmeared angular distribution. Note, however, that the angular 
distribution obtained this way is the distribution for accepted events 
which is different from the
$1+P(W)\alpha_1 \cos\theta^{*} + \alpha_{2} \cos^{2}\theta^{*}\:$
distribution expected for all events. Since the correlation between the angle
and the transverse mass depends on the transverse
momentum  of the $W$ boson, a separate correlation plot is used
for each \pt\ bin. The correlation does not depend on the angular parameters
$\alpha_{1}$, and  $\alpha_{2}$, however, as can be seen from equation
~\ref{eqn:corr}~\cite{ma}:
\begin{equation}
m_{T}^W  =  \frac{m_{e\nu}}{\sqrt{2}} \times
 \sqrt{2 \sqrt{a_{0}+a_{1}\gamma^{2} +a_{2}\gamma^{4}} 
  \mbox{} -2(-\sin^{2}\theta^{*}
 \mbox{} +\gamma^{2}
(1-\cos^{2}\phi^{*}
\sin^{2}\theta^{*}
)}
\label{eqn:corr}
\end{equation}
where $m_{e\nu}$ is the invariant mass of the $e\nu$ system and  the various 
parameters are defined as:
\begin{eqnarray*}
\gamma&=&\frac{p_{T}^W}{m_{e\nu}}\;  \\
a_{0}&=&\sin^{4}\theta^{*}\\
a_{1}&=&2\sin^{2}\theta^{*}(\sin^{2}\phi^{*}-\cos^{2}\phi^{*}\cos^{2}\theta^{*})\\
a_{2}&=&(1-\cos^{2}\phi^{*}\cos^{2}\theta^{*})^{2}\\
\end{eqnarray*}
This equation provides the analytical expression 
for the dependence of the transverse mass on the angles $\theta^{*}$ and
$\phi^{*}$, where $\phi^{*}\:$ is the azimuthal angle in the
Collins-Soper frame with respect to the $x$-axis:~\cite{ma}. 
Note that in this analysis we integrate over $\phi^*$.

\noindent To obtain an angular distribution from a measured transverse mass
distribution,
the transverse mass distribution has to be inverted using the following
 probability function:
\begin{equation}
f(\cos\theta ^{*}|m_{T})=\frac{g(m_{T}| \cos\theta ^{*})h(\cos\theta ^{*})}{\int g(m_{T}| \cos\theta ^{*})h(\cos\theta ^{*}) d \cos\theta ^{*}} 
\end{equation}
%
where:
\begin{itemize}
\item{$g(m_{T}| \cos\theta ^{*}) $ is  the probability of measuring $m_T$ 
given a certain  $ \cos\theta ^{*}$ value (obtained from Monte Carlo)
;}
\item{$h(\cos\theta ^{*})$ 
is the prior probability
for 
$ \cos\theta ^{*}$:
$ (1+ \cos^{2} \theta^{*})$. It reflects all prior 
knowledge, i.e. the angular distribution in the absence of QCD effects;}

\item{$f(\cos\theta^{*}|m_{T})$ is the probability that an event with transverse mass $m_T^W$ has
a decay angle $\cos \theta ^*$. }
\end{itemize}

It is important to note that
$g(m_{T}| \cos\theta ^{*})=\frac{N_{i,j,accepted}}{\sum_{j}{N_{i,j,all}}}$
 where $i\:$ is the bin number in $ m_{T}\: $ and
$j\:$ is the bin number in $ \cos\theta^{*}$.  
The resulting  $\cos\theta^{*}$ distribution is then the angular
distribution for accepted events.

The angular distribution can now be inferred from the measured $m_{T}^W$
distribution by integrating $f(\cos\theta ^{*}|m_{T})$ over $m_T^W$:

\vskip -0.5cm
\begin{equation}
 N_{j}=\sum_{i}^{all\, m_T\, bins}N_{i}^{m_T}f(\cos\theta ^{*}_{j}|m_{Ti})
\end{equation}
where
$N_{j}\:$ is the number of events in $\cos\theta^{*}$ bin $j$.
%

\begin{figure}[!htbp]
\vbox{
\centerline{
\epsfig{figure=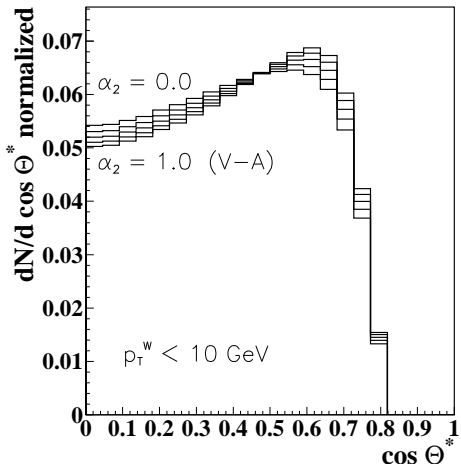,height=7cm,width=7cm}}
\caption{Templates of the angular distribution for various $\alpha_{2}$ 
values for $p_T^W \leq$ 10 GeV. These templates are obtained from Monte
Carlo after acceptance cuts have been applied which results in the drop-off at small angles.}
\label{fig:temp}
}
\end{figure}

To measure the angular parameter $\alpha_{2}$ a series of $\cos\theta ^*$ templates is  
generated from Monte Carlo in bins of $p_T^W$, each 
normalized to unity. Each of the templates is generated taking the 
transverse mass distribution from a high statistics Monte Carlo sample 
for a specific $\alpha_{2}\:$value and converting it into an angular 
distribution by means of the Bayesian method described previously.
Detector effects are included by applying smearing, efficiency, and
acceptance corrections in the Monte Carlo program.
The angular distribution obtained from data is  then compared to 
these templates to determine which value for  $\alpha_{2}\:$ fits best.   
Figure ~\ref{fig:temp} shows a series of such templates for 
$p_{T}^W\leq$ 10 GeV.

\subsection{Data Selection}
The $W$ sample has been selected from data taken during the 1994-95 run of the
Tevatron, and corresponds to an integrated luminosity of 
$84.5\;\rm pb^{-1}$. 

The measurements of the $W$ boson angular decay distribution used the
decay mode $W\rightarrow e\nu$.
Electrons were detected in hermetic, uranium/liquid-argon calorimeters
with an energy resolution of about $15\%/\sqrt{E (\mbox{GeV})}$.
The calorimeters have a granularity of $\Delta\eta \times
\Delta\phi = 0.1 \times 0.1$, where $\eta$ is the pseudorapidity
and $\phi$ is the azimuthal angle.
Electrons were accepted in the region $|\eta|<1.1$ (central calorimeter).
We assume that
the transverse momentum of the neutrino is given by
the calorimetric measurement of the missing
transverse energy ({\mbox{$\not\!\!E_T$}}) in the event.

Electrons from $W$ and $Z$ boson decays tend to be isolated.
Thus, we made the cut
$$\frac{E_{tot}(0.4)-E_{EM}(0.2)}{E_{EM}(0.2)}<0.15,$$
where $E_{tot}(0.4)$ is the energy within $\Delta R < 0.4$ of the cluster
centroid ($\Delta R = \sqrt{\Delta \eta^2 + \Delta \phi^2}$) and $E_{EM}(0.2)$
is the energy in the EM calorimeter within $\Delta R < 0.2$.

At trigger level, a single electron with
$E_T$ greater than 20 GeV was required.
Off-line, a tighter requirement on the electron quality was introduced to reduce the
background level from QCD dijet events, especially at high transverse momentum~\cite{miguel}.
Electron identification was based on a likelihood technique.
Candidates were first identified by finding isolated clusters of energy in the EM 
calorimeter with a matching track in the central detector.
We then cut on a likelihood constructed from the following four variables:
the $\chi^2$ from a covariance matrix which measures the consistency of the calorimeter
cluster shape with that of an electron shower;
the electromagnetic energy fraction (defined as the ratio of the 
energy of the cluster found in the EM calorimeter to its total energy); 
a measure of the consistency between the track position and the cluster centroid;
and the ionization $dE/dx$ along the track.
We
require one central isolated electron and \met$>25$ GeV.
The event is rejected if there is a second electron and the 
dielectron invariant mass lies in the range $75-105$ GeV.
A total of 41173 central $W$ candidates passed these cuts.
A parametric Monte Carlo program~\cite{wmassprd}
was used to simulate the \D0 detector response and
calculate the kinematic and geometric acceptance
as a function of $m_T$ and $p_T$.
The detector resolutions used in the
Monte Carlo were determined from data, and were parameterized as
a function of energy and angle. The relative response of the hadronic and EM
calorimeters was studied using the transverse momentum of the $Z$ boson
as measured by the $p_T$ of the two electrons compared to the hadronic recoil
system in the $Z$ event.
\subsection {Backgrounds}
In order to use the transverse mass distribution of $W$ Bosons to measure the 
angular distribution, the size of the backgrounds and
their dependence on the two variables of interest here, transverse 
momentum and transverse mass, have to be estimated. The following sections describe how the four 
dominant backgrounds are calculated and how  they depend on 
transverse mass and momentum. 
\subsubsection{QCD}
\label{QCDBKG}
QCD dijet events in which a
jet is misidentified as an electron and the energy in the event is mismeasured 
which results in large missing transverse energy pose the largest
overall background. 
The reason for this is the very large dijet cross section compared to
the $W$ cross section. This background is estimated from a data sample
that is dominated by fake electrons which can be obtained by 
replacing
the  \met  cut for W events with a complementary cut 
\met $< 15$ GeV which selects events with fake electrons from multijet events  with only
a small contamination from real W events.
The overall QCD background fraction is $f_{QCD}^W=(0.95 \pm 0.6)\%\:$ 
($f_{QCD}^W=(0.77 \pm 0.6)\%\:$) without (with) a transverse mass cut of 
$50 < m_T^W < 90$ GeV  imposed, 
which is the range used in the Bayesian analysis. 
\subsubsection{$Z\rightarrow ee$}
$Z$ events can look like $W$ events if one electron is lost in an 
uninstrumented or under-instrumented  region of the detector  
and the 
resulting energy imbalance amounts to large \met . This background
can only be estimated using Monte Carlo $Z$ events. The number of such $Z$ events 
present in the $W$ sample is calculated by applying the $W$ selection cuts to 
\texttt{HERWIG}~\cite{herwig} 
$Z\rightarrow ee$ events that were passed through a \texttt{GEANT} based simulation of the \D0 detector  and were  overlaid with 
events from random 
$p$\pbar\ crossings.
The overall background fraction is$f_{Z}^{W} = 0.0091 \pm 0.0013 $ for all \pt .
\subsubsection{Top-Antitop}
The top background is  calculated in a similar way as the $Z$ background from
\texttt{HERWIG} $t$\tbar $\:$ events.
The overall background fraction is$f_{top}^{W} = 0.0016 \pm 0.0005 $ for all 
\pt.
\subsubsection{$W\rightarrow\tau\nu$}
$ W\rightarrow \tau$ events where the $\tau$ subsequently decays into an electron and two 
neutrinos are indistinguishable from $ W\rightarrow e\nu$ events.
This background is included in the parameterized Monte Carlo described above: 
2.22 \% of the events are generated as  $\tau$'s so that the plot 
correlating $\cos\theta^*$ and $m_T^W$ simply takes these  events into account. 

\begin{figure}[!htbp]
\vbox{
\centerline{
\epsfig{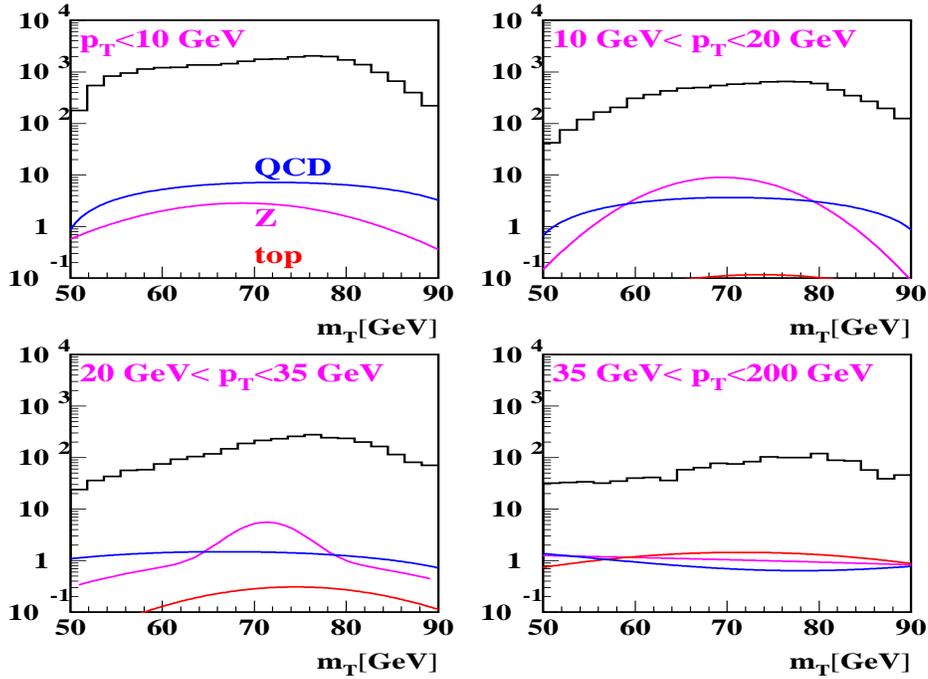}
}
\caption{Summary of all backgrounds in four \pt\ bins.}
\label{fig:allbkg}
}
\end{figure}

Figure~\ref{fig:allbkg} shows the transverse mass distribution of 
$W\rightarrow e\nu $ candidates together with  all backgrounds 
(excluding $\tau$) in four $p_T^W$ bins. All background rates 
are rather small. The background shapes shown here are fitted with heuristic 
functions and the errors on the fits are used to determine the systematic errors 
in the measurement of $\alpha_2$ due to background subtraction.
%
%
%
%
%
\subsection {The measurement of $\alpha_{2}$} 
\begin{figure}[!htbp]
\vbox{
\centerline{
\epsfig{figure=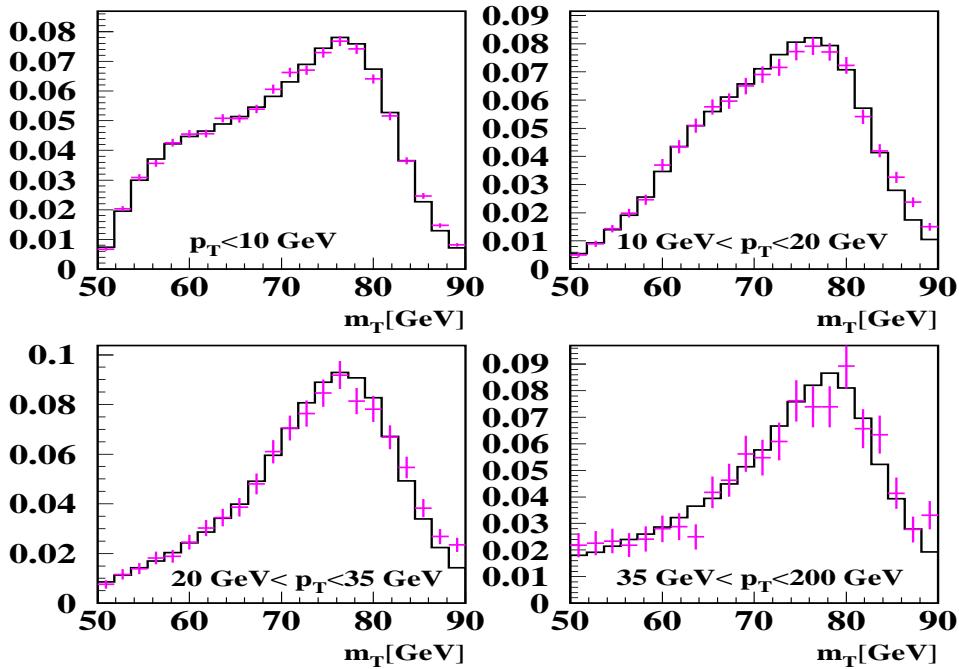,height=4in,width=5.5in}}
\caption{Background subtracted transverse mass distributions(crosses) in four \pt bins
compared to Monte Carlo predictions (lines). (\D0 preliminary)}
\label{fig:fig8}
}
\end{figure}

To obtain the angular distribution for $W$ events from data, the transverse 
mass distribution is inverted according to the Bayesian method 
described in  Section~\ref{extraction}. Figure~\ref{fig:fig8} shows the background 
subtracted transverse mass distributions for the four $p_{T}\:$ bins 
used in this 
analysis. The $\chi^{2}$-values per degree of freedom are
1.67, 1.77, 1.02, and 1.10, respectively, in order of increasing \pt .

To extract the angular parameter $\alpha_{2}$ from the angular distribution
obtained by inverting the transverse mass distribution a log-likelihood 
method is used,\\
\begin{eqnarray}
\ln {\mathcal L} = \sum _{i}^{all\, \cos\theta^*\,bins} n_i \ln p_i
\end{eqnarray}
where $p_{i}$ is the normalized population of a $ \cos\theta^{*}\:$ bin
for one of the Monte Carlo templates and $n_{i}$ is the population
of the same bin in the angular distribution obtained from data.
The statistical errors for $\alpha_{2}\:$ are taken at the points where 
$\ln \mathcal{L}\:$ drops by 0.5 units.

\begin{figure}[!htbp]
\vbox{
%
\centerline{
\epsfig{figure=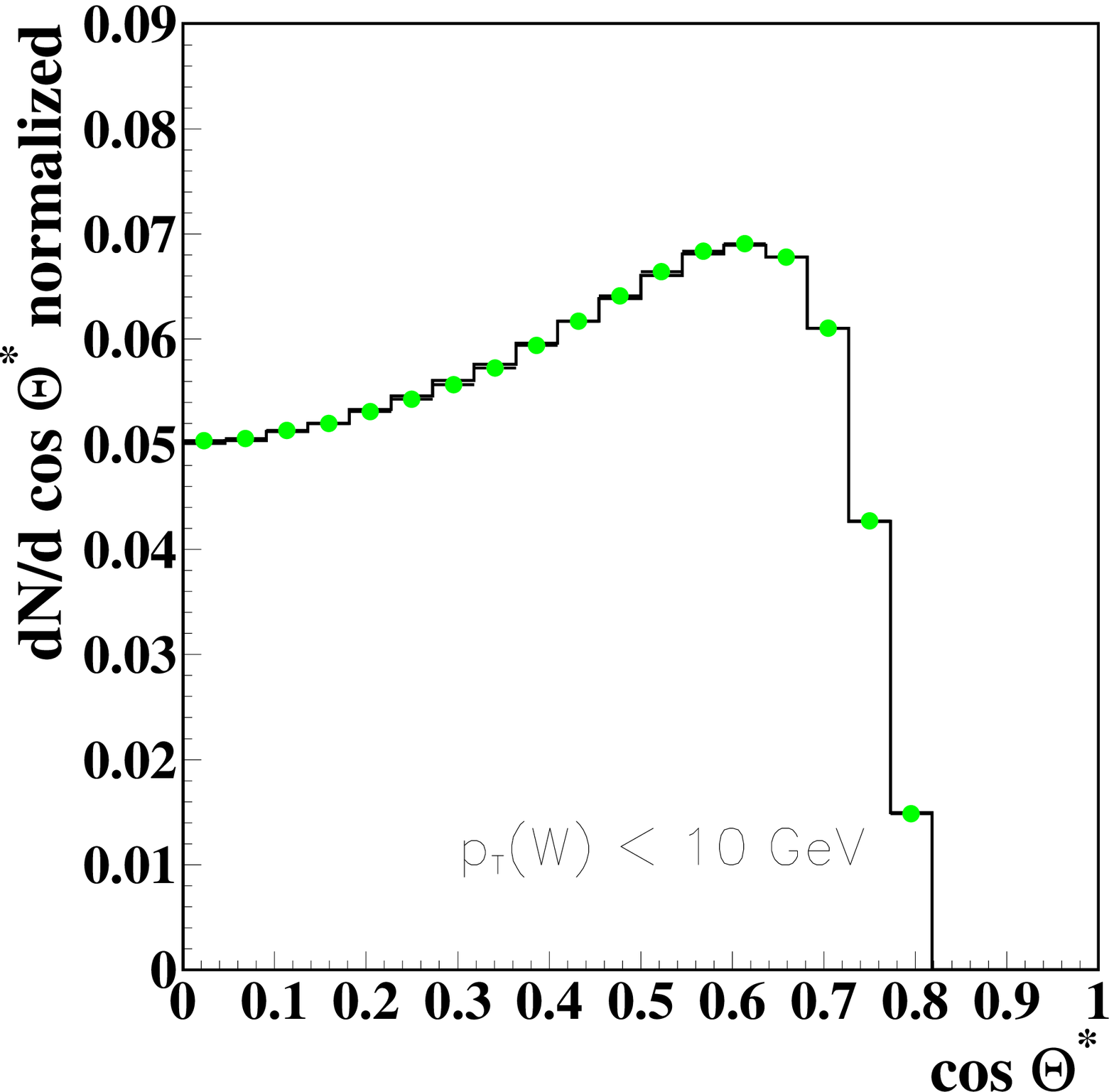,height=2in,width=2.8in}
\epsfig{figure=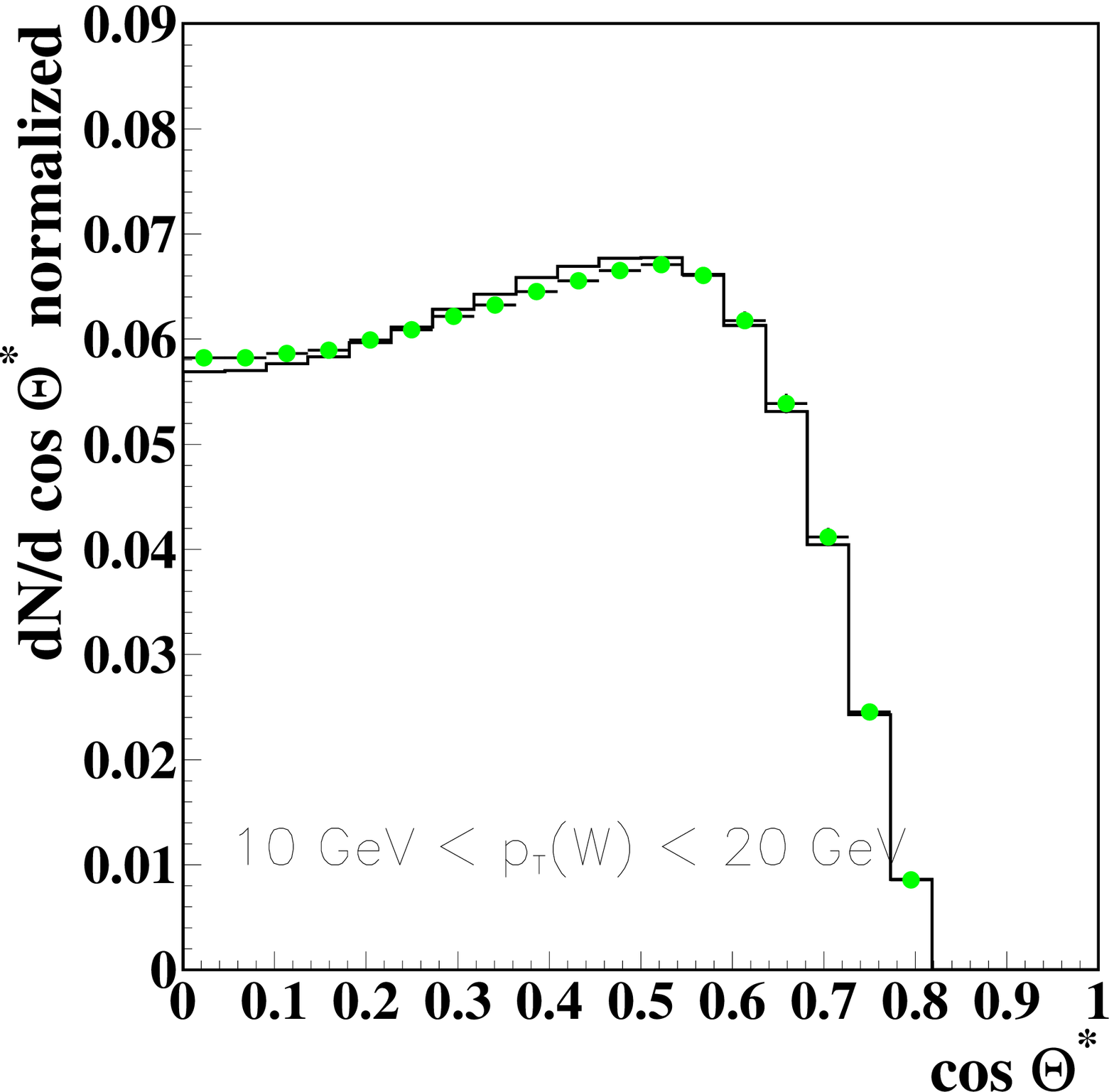,height=2in,width=2.8in}
}
\centerline{
\epsfig{figure=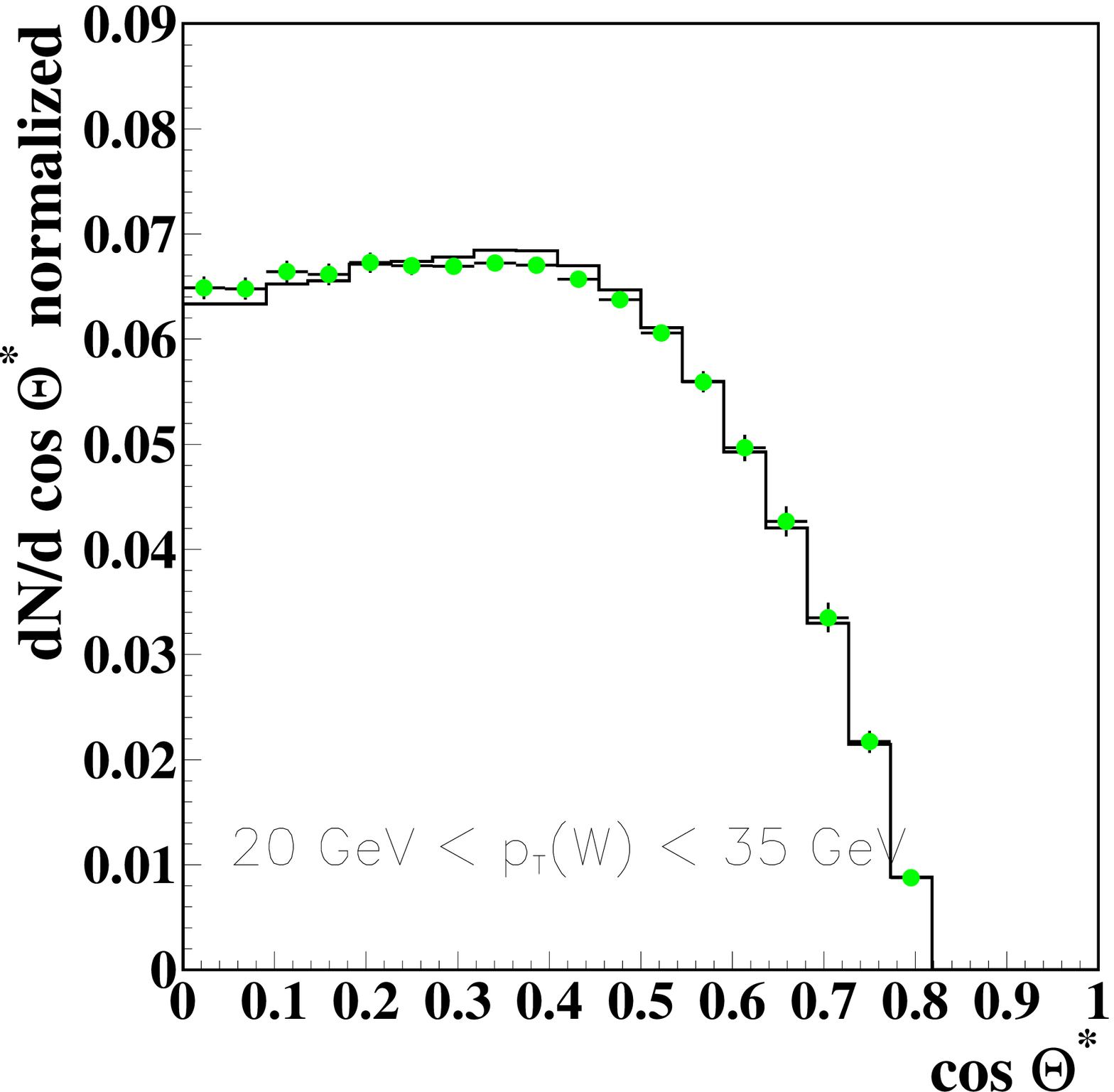,height=2in,width=2.8in}
\epsfig{figure=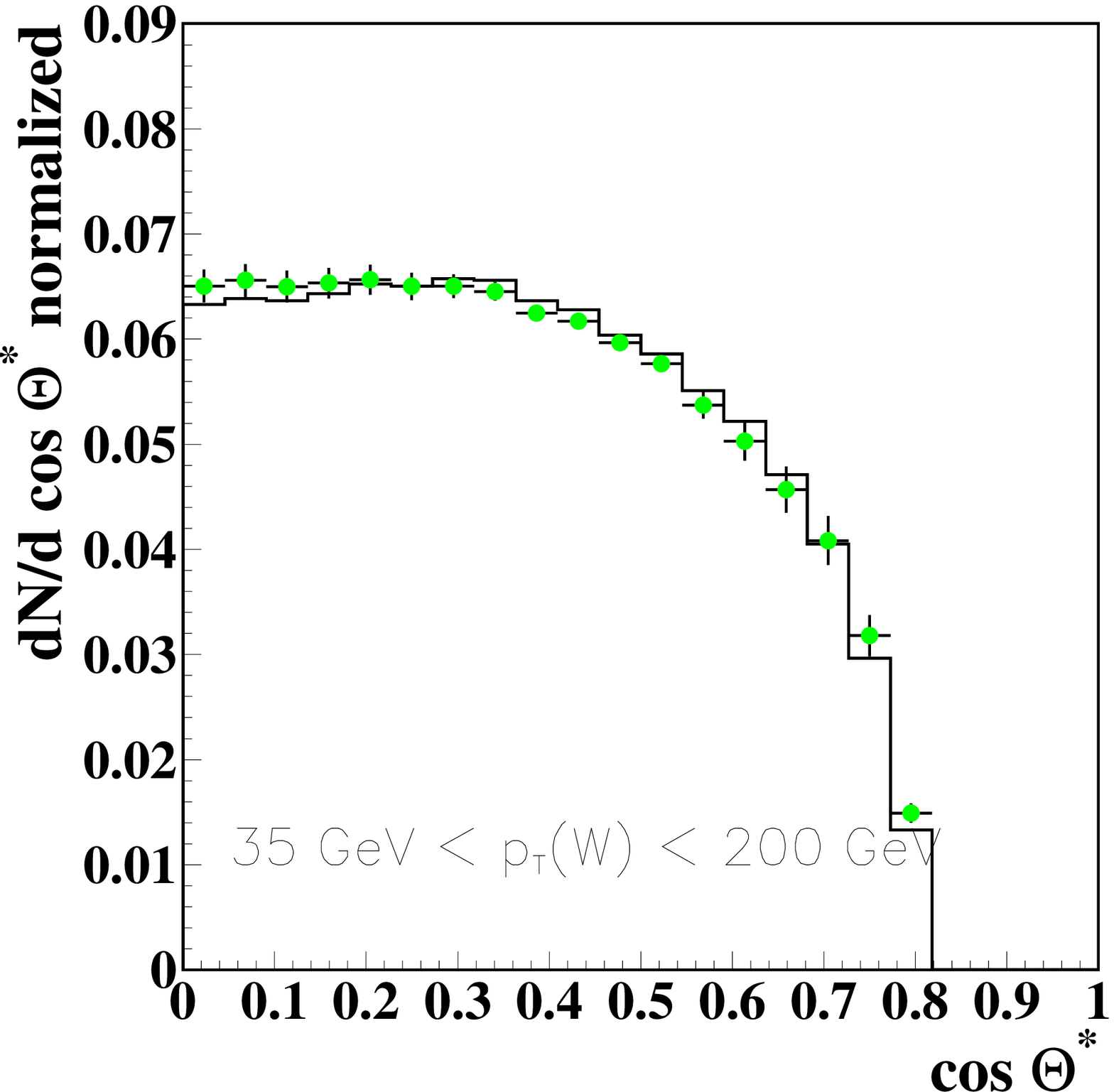,height=2in,width=2.8in}
}
\vspace{0.4cm}
\caption{Angular distributions for data (crosses) compared to Monte Carlo templates (lines) for four different \pt bins. (\D0 preliminary)}
\label{fig:compang}
}
\end{figure}

\begin{figure}[!htbp]
\vbox{
%
\centerline{
\epsfig{figure=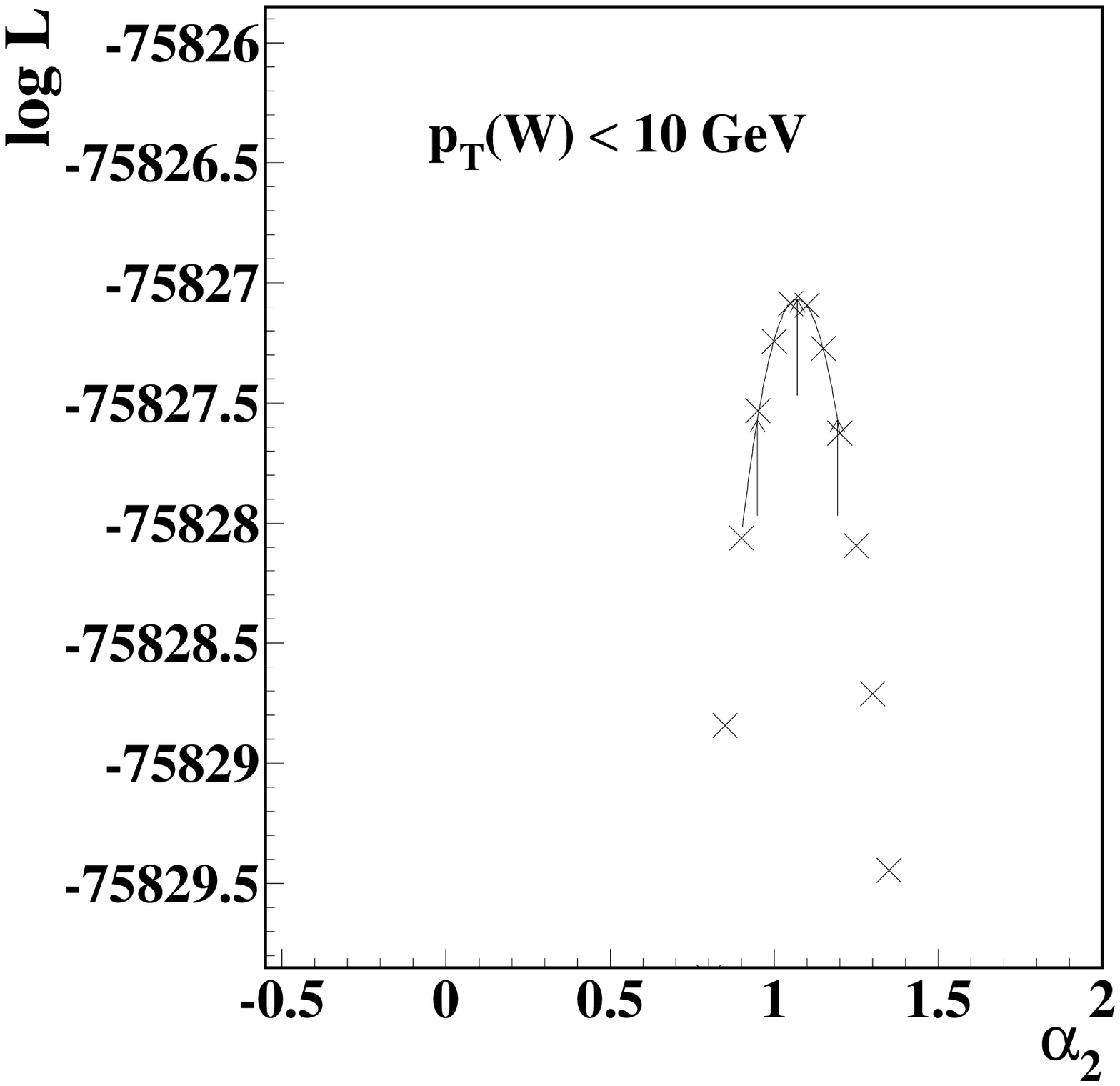,height=2.in,width=2.8in}
\epsfig{figure=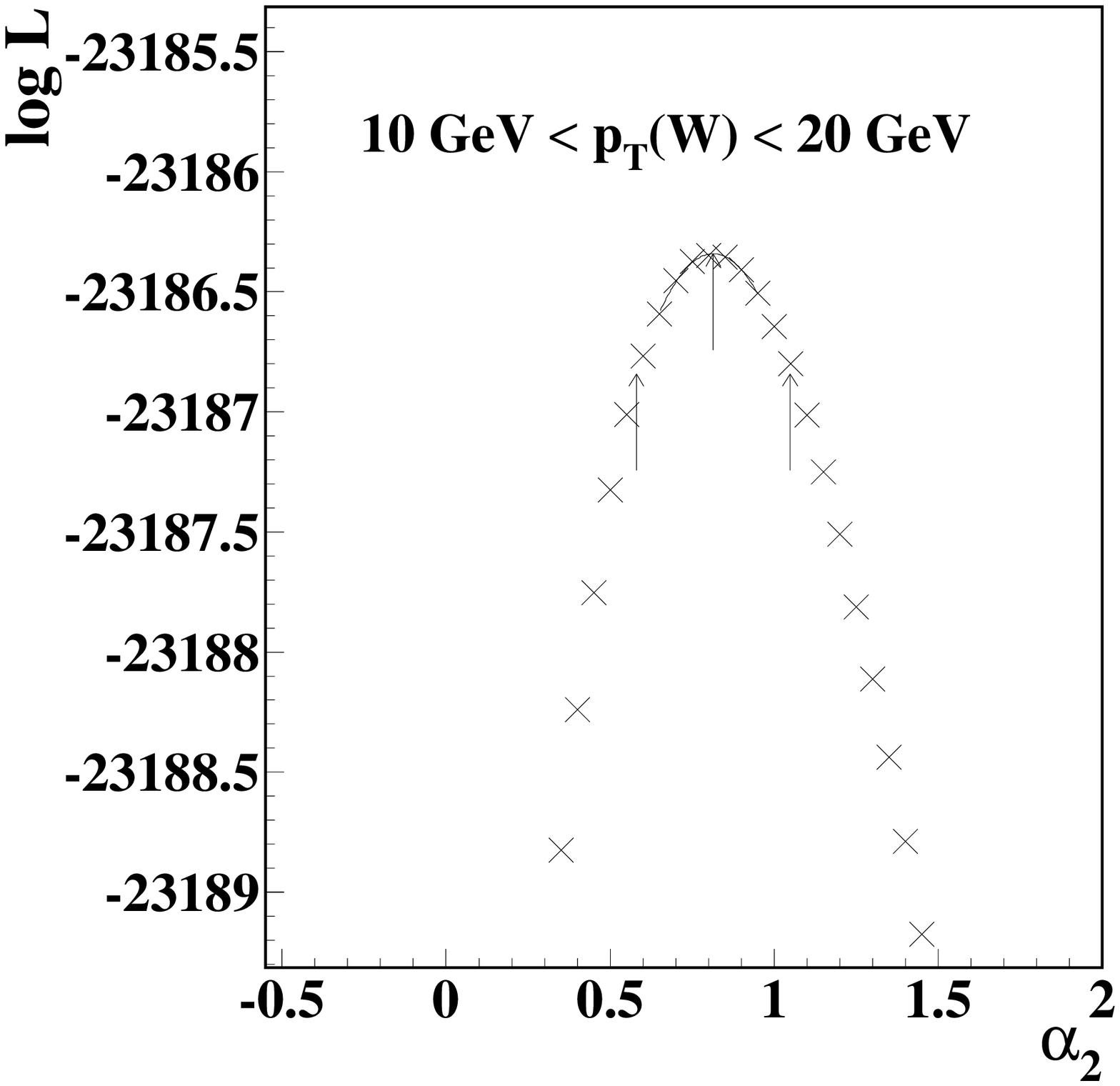,height=2.in,width=2.8in}
}
\centerline{
\epsfig{figure=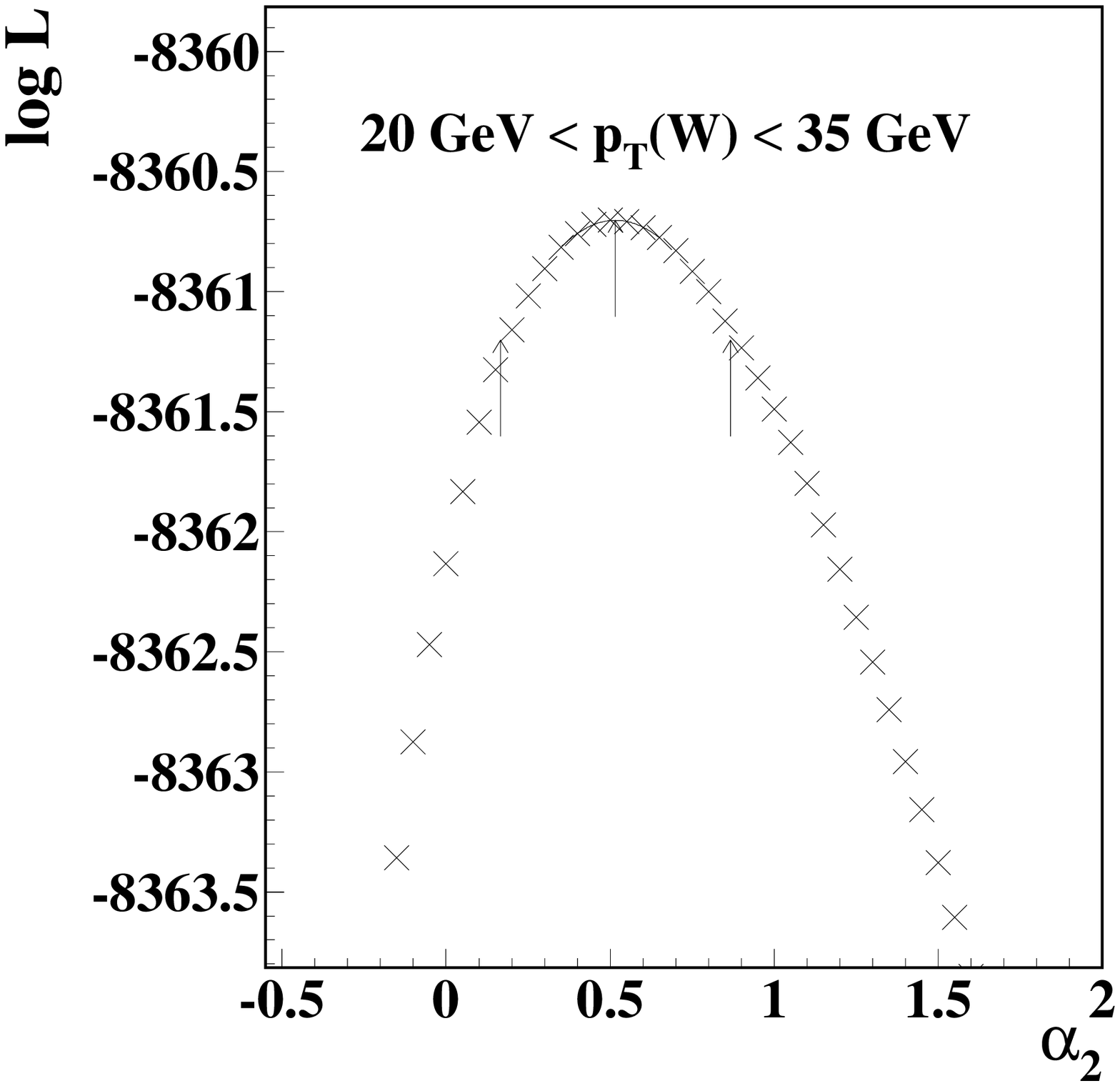,height=2.in,width=2.8in}
\epsfig{figure=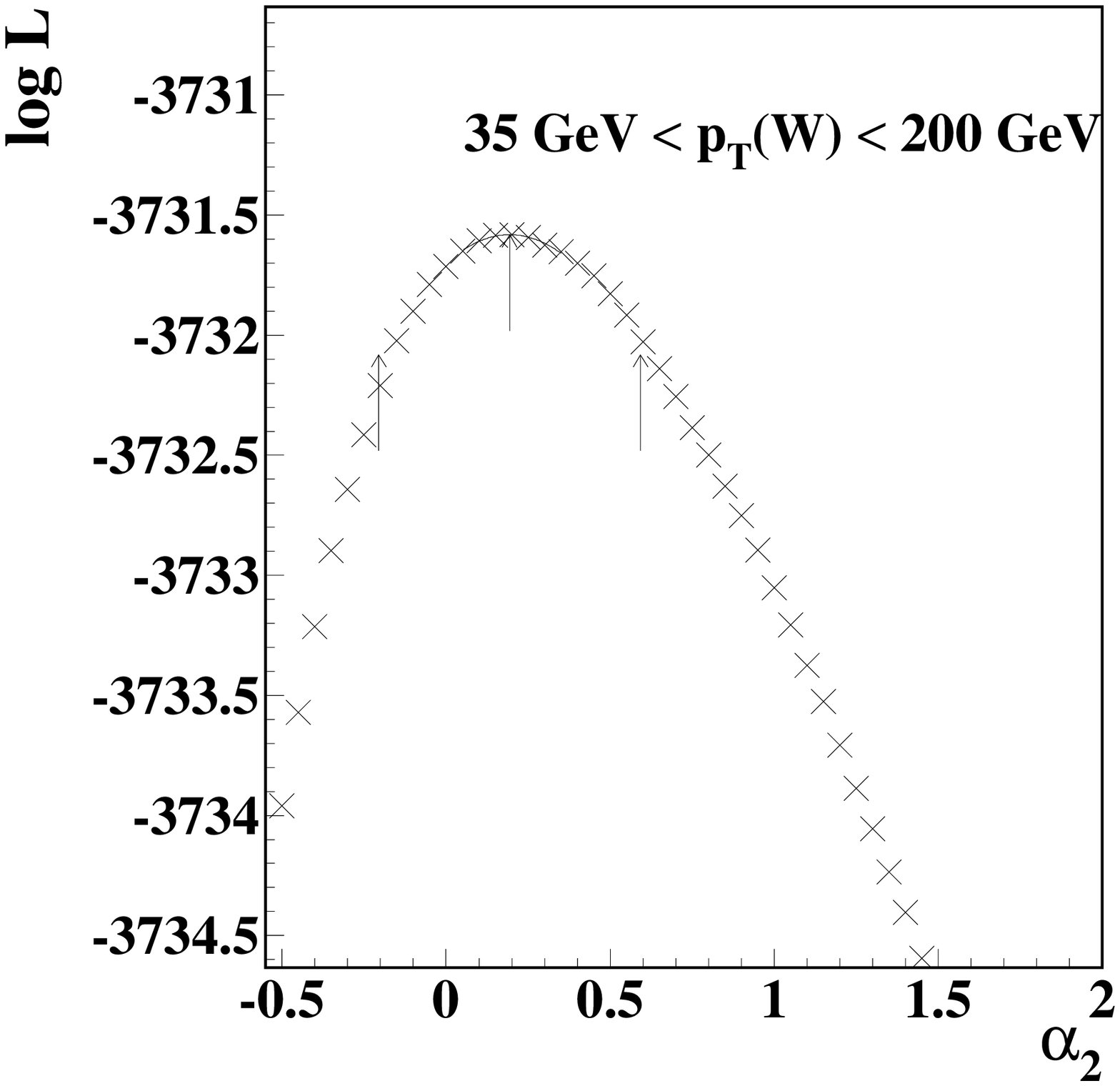,height=2.in,width=2.8in}
}
\caption{Log likelihood functions for four different \pt bins.}
\label{fig:logld}
}
\end{figure}

\begin{figure}[!htbp]
\vbox{
\centerline{
\epsfig{figure=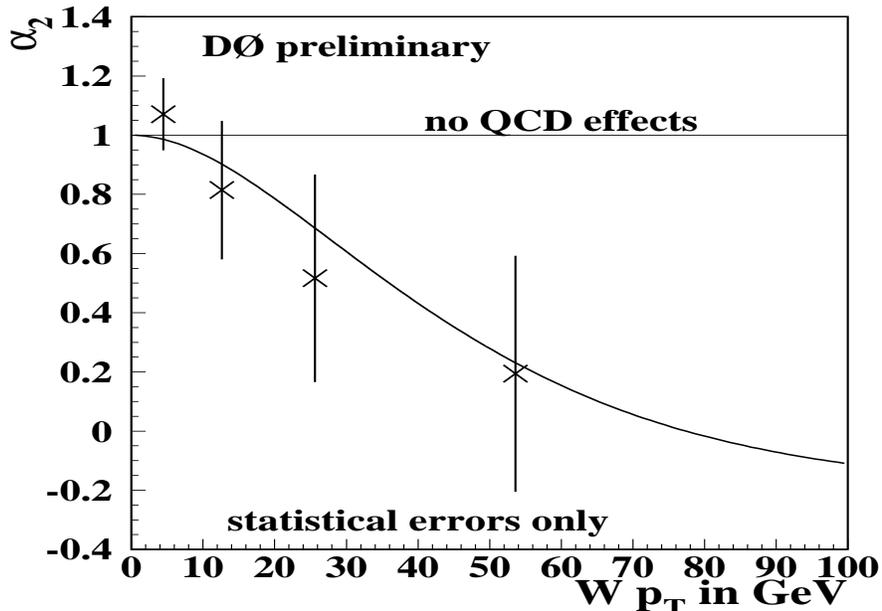,height=3.6in,width=5in}
}
\caption{Measured $\alpha_{2}$ as a function of \pt and its statistical errors compared 
to next-to-leading order QCD calculation by Mirkes (curve) and calculation in the 
absence of QCD (horizontal line).}
\label{fig:result}
}
\end{figure}

In figure~\ref{fig:compang} the angular distributions obtained from data are compared
to the Monte-Carlo templates that fit best.  
In figure~\ref{fig:logld} the log-likelihood distributions 
for  $\alpha_{2}\:$ are shown in the four $p_{T}$ ranges.
To estimate the sensitivity of this experiment the $\chi^{2}\:$ of the 
$\alpha_{2}$ distribution is calculated with respect to the next-to-leading
order QCD prediction and with respect to $(V-A)$ theory in the absence of QCD.
%
The  $\chi^{2}\:$  with respect to the QCD prediction is 0.9/4 dof  which 
corresponds to a probability of $93\%$.
The  $\chi^{2}\:$ with respect to the no-QCD prediction is 7.0/4 dof  which 
corresponds to $14 \%$. Using the odds-ratio technique, the next-to-leading 
order QCD calculation is found to be preferred by   $\approx 2 \sigma$ over the 
calculation in the absence of QCD.
The results of this measurement, including the dominant sources of error, 
are summarized in figure~\ref{fig:result}
and table~\ref{tab:results}. The systematic errors are estimated by 
varying the relevant parameter (like background shape, overall background 
rate, or modeling parameters) in the Monte Carlo by their errors and 
rerunning the analysis program resulting in a varied angular parameter 
$\alpha_2$. 

\nopagebreak

\vskip -2.0cm
\nopagebreak
\vskip 2cm 
%
%
%
%
%
%
\begin{table}[!htbp]
\begin{tabular}{||l|l|l|l|l||}	\hline
\pt     &   $0 \leq p_{T} \leq 10$ &   $10 \leq p_{T} \leq 20$ &   $20 \leq p_{T} \leq 35$  &   $35 \leq p_{T} \leq 200$ \\ \hline
$\alpha_2$&    $1.07\pm0.12$   & $ 0.81 \pm0.23 $  & $ 0.52 \pm0.35 $  & $ 0.19 \pm0.40 $    \\ \hline
$\alpha_2,\: predicted$&    $0.99$   & $ 0.90$  & $ 0.69$  & $ 0.23 $    \\ \hline
mean \pt&    4.4    &12.6  &25.7  & 53.6         \\ \hline
error QCD & $\pm0.022$ & $\pm0.048$ & $\pm0.086$&  $\pm0.069$        \\ \hline
error QCD shape  & $\pm0.035$ & $\pm0.004$ & $\pm0.026$&  $\pm0.008$        \\ \hline 
error $Z$&    $\pm0.002$ & $\pm0.01$ & $\pm0.01$&  $\pm0.04$        \\ \hline
error $Z$ shape&   $\pm0.00$ & $\pm0.01$ & $\pm0.01$&  $\pm0.02$        \\ \hline 
error top&     $\pm0.0005$ & $\pm0.0002$ & $\pm0.001$&  $\pm0.008$        \\ \hline
error hadronic response  & $\pm0.013$ & $\pm0.00$ & $\pm0.06$&  $\pm0.07$        \\ \hline
error hadronic recoil model&     $\pm0.04$ & $\pm0.09$ & $\pm0.10$&  $\pm0.10$        \\ \hline
\end{tabular}
\vskip+0.5cm
\caption{Central values and statistical errors for $\alpha_{2}$ and systematic errors due to backgrounds and the hadronic energy scale and resolution.}
\label{tab:results}
\end{table}
\section{Conclusions}
Using data taken with the \D0 detector during the 1994--1995 Tevatron collider
run, we have presented the measurement of the 
angular distribution of decay electrons from $W$ boson events.  
A next-to-leading order QCD calculation is preferred by $\approx 2 \sigma$ 
over a calculation where no QCD effects are taken into account. 

\section{Acknowledgements}

\input{acknowledgement_paragraph}


\end{document}

%% file: list_of_authors
%
B.~Abbott,$^{45}$                                                             
M.~Abolins,$^{42}$                                                            
V.~Abramov,$^{18}$                                                            
B.S.~Acharya,$^{11}$                                                          
I.~Adam,$^{44}$                                                               
D.L.~Adams,$^{54}$                                                            
M.~Adams,$^{28}$                                                              
S.~Ahn,$^{27}$                                                                
V.~Akimov,$^{16}$                                                             
G.A.~Alves,$^{2}$
% ...
\\                                                                            
\vskip 0.30cm                                                                 
\centerline{(D\O\ Collaboration)}                                             
\vskip 0.30cm                                                                 
\centerline{$^{1}$Universidad de Buenos Aires, Buenos Aires, Argentina}       
\centerline{$^{2}$LAFEX, Centro Brasileiro de Pesquisas F{\'\i}sicas,         
                  Rio de Janeiro, Brazil}                                     
\centerline{$^{3}$Universidade do Estado do Rio de Janeiro,                   
                  Rio de Janeiro, Brazil}                                     
\centerline{$^{4}$Institute of High Energy Physics, Beijing,                  
                  People's Republic of China}                                 
\centerline{$^{5}$Universidad de los Andes, Bogot\'{a}, Colombia}             
\centerline{$^{6}$Universidad San Francisco de Quito, Quito, Ecuador}         
\centerline{$^{7}$Institut des Sciences Nucl\'eaires, IN2P3-CNRS,             
                  Universite de Grenoble 1, Grenoble, France}                 
\centerline{$^{8}$DAPNIA/Service de Physique des Particules, CEA, Saclay,     
                  France}                                                     
\centerline{$^{9}$Panjab University, Chandigarh, India}                       
\centerline{$^{10}$Delhi University, Delhi, India}                            
\centerline{$^{11}$Tata Institute of Fundamental Research, Mumbai, India}     
\centerline{$^{12}$Kyungsung University, Pusan, Korea}                        
\centerline{$^{13}$Seoul National University, Seoul, Korea}                   
\centerline{$^{14}$CINVESTAV, Mexico City, Mexico}                            
\centerline{$^{15}$Institute of Nuclear Physics, Krak\'ow, Poland}            
\centerline{$^{16}$Institute for Theoretical and Experimental Physics,        
                   Moscow, Russia}                                            
\centerline{$^{17}$Moscow State University, Moscow, Russia}                   
\centerline{$^{18}$Institute for High Energy Physics, Protvino, Russia}       
\centerline{$^{19}$Lancaster University, Lancaster, United Kingdom}           
\centerline{$^{20}$University of Arizona, Tucson, Arizona 85721}              
\centerline{$^{21}$Lawrence Berkeley National Laboratory and University of    
                   California, Berkeley, California 94720}                    
\centerline{$^{22}$University of California, Davis, California 95616}         
\centerline{$^{23}$University of California, Irvine, California 92697}        
\centerline{$^{24}$University of California, Riverside, California 92521}     
\centerline{$^{25}$Florida State University, Tallahassee, Florida 32306}      
\centerline{$^{26}$University of Hawaii, Honolulu, Hawaii 96822}              
\centerline{$^{27}$Fermi National Accelerator Laboratory, Batavia,            
                   Illinois 60510}                                            
\centerline{$^{28}$University of Illinois at Chicago, Chicago,                
                   Illinois 60607}                                            
\centerline{$^{29}$Northern Illinois University, DeKalb, Illinois 60115}      
\centerline{$^{30}$Northwestern University, Evanston, Illinois 60208}         
\centerline{$^{31}$Indiana University, Bloomington, Indiana 47405}            
\centerline{$^{32}$University of Notre Dame, Notre Dame, Indiana 46556}       
\centerline{$^{33}$Purdue University, West Lafayette, Indiana 47907}          
\centerline{$^{34}$Iowa State University, Ames, Iowa 50011}                   
\centerline{$^{35}$University of Kansas, Lawrence, Kansas 66045}              
\centerline{$^{36}$Kansas State University, Manhattan, Kansas 66506}          
\centerline{$^{37}$Louisiana Tech University, Ruston, Louisiana 71272}        
\centerline{$^{38}$University of Maryland, College Park, Maryland 20742}      
\centerline{$^{39}$Boston University, Boston, Massachusetts 02215}            
\centerline{$^{40}$Northeastern University, Boston, Massachusetts 02115}      
\centerline{$^{41}$University of Michigan, Ann Arbor, Michigan 48109}         
\centerline{$^{42}$Michigan State University, East Lansing, Michigan 48824}   
\centerline{$^{43}$University of Nebraska, Lincoln, Nebraska 68588}           
\centerline{$^{44}$Columbia University, New York, New York 10027}             
\centerline{$^{45}$New York University, New York, New York 10003}             
\centerline{$^{46}$University of Rochester, Rochester, New York 14627}        
\centerline{$^{47}$State University of New York, Stony Brook,                 
                   New York 11794}                                            
\centerline{$^{48}$Brookhaven National Laboratory, Upton, New York 11973}     
\centerline{$^{49}$Langston University, Langston, Oklahoma 73050}             
\centerline{$^{50}$University of Oklahoma, Norman, Oklahoma 73019}            
\centerline{$^{51}$Brown University, Providence, Rhode Island 02912}          
\centerline{$^{52}$University of Texas, Arlington, Texas 76019}               
\centerline{$^{53}$Texas A\&M University, College Station, Texas 77843}       
\centerline{$^{54}$Rice University, Houston, Texas 77005}                     

%% file: acknowledgement_paragraph
%
We thank the Fermilab and collaborating institution staffs for
contributions to this work and acknowledge support from the 
Department of Energy and National Science Foundation (USA),  
Commissariat  \` a L'Energie Atomique (France), 
Ministry for Science and Technology and Ministry for Atomic 
   Energy (Russia),
CAPES and CNPq (Brazil),
Departments of Atomic Energy and Science and Education (India),
Colciencias (Colombia),
CONACyT (Mexico),
Ministry of Education and KOSEF (Korea),
and CONICET and UBACyT (Argentina).
%